\documentclass[runningheads]{llncs}

\usepackage{graphicx}
\usepackage{amsmath}
\usepackage{tabularx}
\makeatletter
\renewcommand*{\@fnsymbol}[1]{\ifcase#1\or\else\@arabic{#1}\fi}
\makeatother

\begin{document}

\title{Assessing the Benefits of Model Ensembles in Neural Re-Ranking for Passage Retrieval \thanks{This research was supported by Funda\c{c}\~{a}o para a Ci\^{e}ncia e Tecnologia (FCT), through the Ph.D. scholarship with reference SFRH/BD/150497/2019, and the INESC-ID multi-annual funding from the PIDDAC programme (UIDB/50021/2020).}}
\titlerunning{Model Ensembles for Neural Passage Re-Ranking}
%
\author{Luís Borges\inst{1,2}\and
Bruno Martins\inst{1} \and Jamie Callan\inst{2}}
\authorrunning{Borges et al.}

\institute{Instituto Superior Técnico and INESC-ID, University of Lisbon, Portugal \and
Carnegie Mellon University, USA\\
}
\maketitle              
\begin{abstract}
Our work aimed at experimentally assessing the benefits of model ensembling within the context of neural methods for passage re-ranking. Starting from relatively standard neural models, we use a previous technique named Fast Geometric Ensembling to generate multiple model instances from particular training schedules, then focusing or attention on different types of approaches for combining the results from the multiple model instances (e.g., averaging the ranking scores, using fusion methods from the IR literature, or using supervised learning-to-rank). Tests with the MS-MARCO dataset show that model ensembling can indeed benefit the ranking quality, particularly with supervised learning-to-rank although also with unsupervised rank aggregation.

\end{abstract}
\section{Introduction}
Ensemble methods are known to typically perform better than individual systems. In the field of information retrieval, several rank aggregation techniques have for instance been proposed to combine the results of different ranking methods~\cite{anava2016probabilistic,cormack2009reciprocal,lillis2010estimating,lillis2008extending}, with previous studies showing that ensembles indeed lead to superior results. Ensemble methods are also common in the machine learning literature. Specifically within the context of learning with deep neural networks, ensembling algorithms such as Fast Geometric Ensembling (FGE) have recently been proposed and successfully applied to multiple tasks \cite{garipov2018loss}, using particular learning rate updating schedules to create multiple neural networks with no additional training cost, which can afterwards be combined (e.g., by averaging the scores from the resulting models) for improved performance.

In this paper, we assess the benefits of ensemble approaches within the context of neural models for passage re-ranking. We specifically leverage the FGE approach together with relatively standard neural retrieval models~\cite{lin2020pretrained}, corresponding to re-ranking approaches based on recurrent neural networks, or instead based on Transformer-based models like RoBERTa~\cite{liu2019roberta}. Leveraging model ensembles, we focused our attention on different approaches for combining the results, and we compared strategies based on (a) averaging the scores (i.e., the relevance estimates) produced by the multiple models, (b) combining the rankings from the multiple models with rank fusion approaches, or (c) using supervised learning-to-rank as a meta-learning strategy to combine the model scores.

We evaluated the different approaches on the well-known MS-MARCO passage re-ranking task~\cite{bajaj2016ms}. The obtained results show that model ensembling indeed leads to improvements over individual neural ranking models, particularly with supervised learning-to-rank and/or in the case of RoBERTa models.


\section{Passage Re-Ranking with Neural Ensembles}

\vspace{-0.3cm}
Within our general approach, we first train a neural ranking model with the Fast Geometric Ensembling (FGE) technique, which outputs $N$ different model checkpoints, saved at different phases of the training process. The different checkpoints are used to re-rank initial lists with the top 1000 passages for each test query, resulting in the generation of $N$ ranked lists. For the initial rankings, we used the DeepCT first-stage retrieval algorithm, which extends BM25 with context-aware term weights derived from a BERT model~\cite{dai2019context}. Finally, the $N$ different ranked lists are used as input to a fusion method, which combines the scores to produce a final re-ranked list. The following sub-sections describe the FGE technique and the different fusion methods that were considered.

\vspace{-0.4cm}
\subsection{Fast Geometric Model Ensembling}

Fast Geometric Ensembling (FGE) consists of an ensembling technique for deep neural networks that generates multiple points in the weight space (i.e., multiple model instances, resulting from different checkpoints during training), that share a similar low test error~\cite{garipov2018loss}. The approach is inspired on the observation that the optima for the loss functions being optimized while training neural models are often connected by simple curves, over which the training/test accuracy are nearly constant. FGE uses a training procedure that leverages this geometric intuition, discovering points (i.e., model checkpoints) within the high-accuracy pathways through a particular learning rate update schedule.

The FGE algorithm starts with model weights corresponding to an initial training of the neural network, and resumes the training with a cyclical learning rate defined as follows, where $\alpha_1$ and $\alpha_2$ are the minimum and maximum values for the learning rate, while $\alpha(i)$ represents the learning rate at iteration $i$.
\begin{equation}
    \alpha(i)= 
\begin{cases}
    \left(1-2 \times \mathrm{t}(i) \right) \times \alpha_1 + 2 \times \mathrm{t}(i) \times \alpha_2 & 0 < \mathrm{t}(i) \leq 0.5 \\
    \left(2-2 \times \mathrm{t}(i) \right) \times \alpha_2 + \left(2 \times \mathrm{t}(i)-1 \right) \times \alpha_1 & 0.5 < \mathrm{t}(i) \leq 1
\end{cases}
\end{equation}
Each iteration corresponds to processing one mini-batch. The parameter $\mathrm{t}(i)$ can be defined with basis on the number of iterations $c$ corresponding to a cycle. 

\begin{equation}
    \mathrm{t}(i) = \frac{1}{c} \times \left(\mathrm{mod}(i-1, c) + 1 \right)
\end{equation}
In the middle of each cycle, when the learning rate reaches its minimum value $\alpha_2$, the model weights are collected to form a checkpoint. After training, the checkpoints can be individually evaluated on a test set, and the corresponding results can afterwards by combined to form ensemble predictions.

\vspace{-0.4cm}
\subsection{Rank Fusion Methods}

Multiple methods for fusing lists into a final consensus ranking have been proposed in the information retrieval literature~\cite{anava2016probabilistic}. As a simple approach, one can for instance rank instances according to the average of the scores associated to the different lists. Other approaches often leverage instead the ranking positions.

One example is Reciprocal Rank Fusion~\cite{cormack2009reciprocal}, which is based on summing the multiplicative inverse of the original rankings. Given a set of instances $P$ (i.e., the passages to be retrieved) and multiple rankings $R$ for a given query, the instances can be sorted according to the following score:
\begin{equation}
    \mathrm{RRFscore}(p \in P) = \sum_{\mathrm{r} \in R} \frac{1}{k + \mathrm{r}(p)}
\end{equation}
In Equation 3, $k$ is a smoothing constant often set to the constant value of 60~\cite{cormack2009reciprocal}, and $\mathrm{r}(p)$ is the rank of passage $p$ in the ranked list $\mathrm{r}()$. A simple variation, named MAP Fusion, was proposed by Lillis et al.~\cite{lillis2010estimating} and involves weighting the contribution of each ranked list according its Mean Average Precision (MAP) score, as measured over a held-out set of queries:
\begin{equation}
    \mathrm{MAPFscore}(p \in P) = \sum_{\mathrm{r} \in R} \frac{1 \times \mathrm{MAP}_{\mathrm{r}}}{k + \mathrm{r}(p)}
\end{equation}

Previous studies have also advanced probabilistic data fusion techniques, using training queries to estimate the probability that a resource is relevant to a given query, and leveraging those probabilities in order to create new ranking scores. One of those probabilistic techniques is SlideFuse~\cite{lillis2008extending}, which first estimates the probability that a passage $p$, occurring in position $i$ of a ranked list produced through a procedure $r$, is relevant. This can be computed according to the following equation, where $Q_p$ is the set of training queries for which at least $i$ instances were returned in lists produced through procedure $r$, and where $\mathrm{Rel}(p_i, q)$ is 1 if $p_i$ is relevant to query $q$, and 0 otherwise.
\begin{equation}
    \mathrm{P}(p_i|r) = \frac{\sum_{q \in Q_i} \mathrm{Rel}(p_i, q)}{Q_i}
\end{equation}
The final aggregated score for each document also considers a sliding window around each position of the rankings to be merged:
\begin{equation}
    \mathrm{SlideFscore}(p \in P) = \sum_{r \in R} \mathrm{P}(p_{i,w}|r)
\end{equation}
In the previous equation, $\mathrm{P}(p_{i,w}|r)$ is the probability of relevance of passage $p$ in position $i$, this time considering a window of $w$ documents around each side of $i$. This can be estimated as follows, where the values $a$ and $b$ correspond to the window limits for every position $i$, considering $N$ as the total number of documents for each query.

\begin{equation}
\begin{gathered}
    \mathrm{P}(p_{i,w}|r) = \frac{\sum_{j=a}^{b} \mathrm{P}(p_j|r)}{b - a + 1}, \text{~with~} 
    \\
    a = 
\begin{cases}
    i-w & i-w \geq 0\\
    0 & i-w<0
\end{cases} \text{~~~and~~~} b = 
\begin{cases}
    i+w & i+w < N\\
    N-1 & i+w \geq N
\end{cases}
\end{gathered}
\end{equation}

Variations on SlideFuse, weighting the contribution of individual ranked lists, are also possible. For instance Equation 6 can be adapted in the same way as Equation 4 extends from Equation 3, weighting each system by the corresponding MAP score, and resulting in a MAP SlideFuse approach.

Besides rank aggregation methods we also experimented with a supervised learning-to-rank approach, specifically the LambdaRank~\cite{burges2010ranknet} implementation from the XGBoost\footnote{\url{https://github.com/dmlc/xgboost}} package. In this case, for each training query, we collected the relevant passage and two other passages in the top 1000 list, ranked according to DeepCT. The LambdaRank model was trained on this data, using as features the DeepCT scores plus those from the FGE snapshots, together with the average and standard deviation, and attempting to optimize the MAP metric.

Still on what regards experimental settings, the SlideFuse method considered a window size of 6, and the LambdaRank algorithm used the default parameters from the XGBoost library, except in the choice of MAP as the optimized metric.

\vspace{-0.2cm}
\section{Neural Ranking Models}
\vspace{-0.2cm}

We experimented with two distinct types of neural ranking models, respectively leveraging recurrent neural networks, and Transformer-based language models. 

The first model is inspired on a previous proposal for encoding and matching textual contents~\cite{10.1145/3287763}. A sentence encoder is used to compute fixed-size vector representations for input sequences, leveraging pre-trained FastText~\cite{bojanowski2017enriching} word embeddings together with two layers of bi-directional LSTM units with shortcut connections between them, and a max-pooling operation over the sequence produced by the second bi-LSTM. The query is processed through the aforementioned encoder, which outputs the corresponding representation. In turn, each sentence that composes the passage is also processed through the same encoder, generating a sequence of representations. This sequence of sentence representations is then fed as input to a different encoder, using a similar structure (except for the initial FastText embedding layer) to produce a single fixed-size representation for the passage. The representations for the query and the passage are combined through different operations (i.e., vector concatenation, difference, and element-wise product), and the result is feed into a final feed-forward layer, which outputs the relevance score of the passage towards the query.


For the second neural ranking approach, we fine-tune RoBERTa-base~\cite{liu2019roberta} to our ranking problem, passing as input to the model the concatenation of the query and the passage text, separated by a special {\tt [SEP]} token. We concatenate the vector representation of the special {\tt [CLS]} token, together with the result of a max-pooling operation over the last sequence of hidden states output by RoBERTa-base, feeding the result to a final feed-forward layer which outputs the relevance score of the passage towards the query.

When training our models, we first use a fast approach (i.e., DeepCT~\cite{dai2019context}) to retrieve the top 1000 passages for the provided training queries. The loss function takes as input the scores between a query and a relevant passage, a non-relevant passage sampled from the top 25 passages retrieved for the query, and a negative passage sampled from the remaining 975 passages in the top 1000. The loss is formally defined as follows, where $p$ is the score between the query and a positive passage, $n_{25}$ is the score between the query and the passage sampled from the top 25, and $n_{975}$ is the score between the query and the passage sampled from the remaining 975 passages. 
\begin{equation}
\begin{gathered}
\mathrm{loss} = \mathrm{hinge}(p, n_{25}) + \mathrm{hinge}(p, n_{975}) + 0.25 \times \mathrm{hinge}(n_{25}, n_{975}) \text{, with} \\
\mathrm{hinge}(p, n) = \max(0, 1 - p + n)
\end{gathered}
\end{equation}

For our RNN-based model, we used a dimensionality of 300 in the representations produced by the recurrent units. For RoBERTa-base, we used the default base parameters as defined in the Huggingface Transformers library\footnote{https://github.com/huggingface/transformers}. We trained our models for a total of 15 epochs with the AdaMod~\cite{ding2019adaptive} optimizer. The first five epochs produced the initial weights for the Fast Geometric Ensembling (FGE) technique. In the remaining ten epochs with FGE, we used cycles of $c=4$ epochs, with a cyclic learning rate between $\alpha_1=2 \cdot 10^{-5}$ and $\alpha_2=2 \cdot 10^{-7}$, hence generating five different checkpoints. 

\vspace{-0.4cm}
\section{Experimental Evaluation}
\vspace{-0.2cm}

Our experiments relied on the passage ranking data from MS-MARCO~\cite{bajaj2016ms}. For each test query, a first-stage ranker (in our case, DeepCT~\cite{dai2019context}) retrieves a set of possibly relevant passages from the whole collection, and the top $k$ results are then re-ranked through a second more expensive model. 

Table~\ref{tab:results} presents a comparison between the different alternatives described in Sections 2 and 3, with results measured over the development portion of the MS-MARCO dataset. We specifically measured the Mean Average Precision (MAP), Mean Reciprocal Rank (MRR), and MRR@10. The first two lines of Table \ref{tab:results} compare two first-stage retrieval approaches, returning 1000 possibly relevant passages for each development query. DeepCT outperformed BM25 in this initial task, and the remaining experiments focused on re-ranking the top 100 passages retrieved by DeepCT. A separate round of tests, not detailed in this paper, showed that re-ranking the top 100 passages lead to consistently better results than re-ranking the entire set of 1000 passages per query.

The second group of rows in Table \ref{tab:results} compares the results for both types of neural models, trained for a total of 15 epochs. The model based on RoBERTa-base clearly outperformed the RNN-based model, which even failed to outperform DeepCT. We also attempted to combine the rankings from each of these models and DeepCT, through the MAPFuse strategy. The results, given in the third group of rows, showed that the combination improved results for the RNN model, but not for the RoBERTa-base model. 

The remaining rows from Table 1 show the results achieved with FGE ensembles, leveraging different types of techniques for combining the rankings. The results show that model ensembling has clear benefits for RoBERTa-base models, with mixed results for RNN models. Few differences were measured between the alternative rank aggregation approaches, and significantly better results were obtained with learning-to-rank. We expect that similar benefits from ensembling can be expected for larger models than RoBERTa-base. 


\section{Conclusions and Future Work}

\begin{table}[t!]
  \begin{center}
    \caption{Results over the MS-MARCO development dataset. Statistical significance tests were used to compare ensembles against individual models for re-ranking the DeepCT results, both for RNN (\dag) and RoBERTa-base (\ddag) models, as well as to compare the learning-to-rank ensembles against the second best ensemble (\textasteriskcentered). The methods whose difference is statistically significant, for a $p$-value of 0.05, are marked on the table. Although this is not reported on the table, not including DeepCT scores in the FGE ensembles is consistently worse (i.e., approx. 0.01 points lower in terms of MRR@10 for RoBERTa-base ensembles, and up to 0.1 points lower for RNN ensembles).}
    \label{tab:results}
    \vspace{-1.0em}
    \scriptsize
    \begin{tabular}{l  c  c  c}
    
       Method &  ~~~~MAP~~~~ &  ~~~~MRR~~~~ &  ~~MRR@10~~\\
       \hline
       BM25 & 0.1835 & 0.1867 & 0.1758\\
       DeepCT & 0.2506 & 0.2546 & 0.2425\\
       \hline
       RNN & 0.2127 & 0.2160 & 0.2010\\
       RoBERTa-base & 0.3356 & 0.3403 & 0.3311\\
       \hline
       RNN + DeepCT & 0.2888 & 0.2936 & 0.2821\\
       RoBERTa-base + DeepCT & 0.3326 & 0.3378 & 0.3285\\
       \hline
       RNN FGE + DeepCT + Average$^\dag$ & 0.3000 & 0.3056 & 0.2952\\
       RNN FGE + DeepCT + RRFuse & 0.2845 & 0.2891 & 0.2769\\
       RNN FGE + DeepCT + MAPFuse & 0.2847 & 0.2893 & 0.2771\\
       RNN FGE + DeepCT + SlideFuse &  0.2738 & 0.2781 & 0.2645\\
       RNN FGE + DeepCT + MAPSlideFuse$^\dag$~~~~~ & 0.2741 & 0.2784 & 0.2649\\
       RNN FGE + DeepCT + Learning-to-Rank$^{\dag*}$ & 0.3131 & 0.3181 & 0.3080\\
       \hline
       RoBERTa-base FGE + DeepCT + Average$^\ddag$ & 0.3354 & 0.3411 & 0.3324\\
       RoBERTa-base FGE +  DeepCT + RRFuse$^\ddag$ & 0.3819 & 0.3879 & 0.3813\\
       RoBERTa-base FGE +  DeepCT + MAPFuse$^\ddag$ & 0.3818 & 0.3874 & 0.3806\\
       RoBERTa-base FGE +  DeepCT + SlideFuse$^\ddag$ & 0.3787 & 0.3844 & 0.3774\\
       RoBERTa-base FGE +  DeepCT + MAPSlideFuse$^\ddag$~~~~~~~~~~~~~~ & 0.3789 & 0.3844 & 0.3774\\
       RoBERTa-base FGE +  DeepCT + Learning-to-Rank$^{\ddag*}$ & 0.3856 & 0.3913 & 0.3846\\
       \hline
    \end{tabular}
  \end{center}
  \vspace{-2.0em}
\end{table}

We tested the use of Fast Geometric Ensembling (FGE) with neural passage re-ranking models, comparing different fusion methods to combine the rankings from FGE checkpoints. Results over MS-MARCO show that model ensembling indeed leads to consistent improvements over individual models, thus constituting a viable approach to further improve state-of-the-art approaches.

For future work, we plan to conduct similar tests with other datasets, including TREC CAR~\cite{dietz2018trec} and WikiPassageQA~\cite{cohen2018wikipassageqa}, in addition to testing different ensembling methods, such as the Auto-Ensembling approach from Jun et al.~\cite{Jun2020}. 

%

%
%
%
 \bibliographystyle{splncs04}
 \bibliography{samplepaper}
\end{document}